\documentclass{article}

\usepackage{amsmath}

\newtheorem{definition}{Definition}[section]
\numberwithin{equation}{section}

\begin{document}

\centerline{How the $\mu$-deformed Segal-Bargmann space gets two measures}

\vskip .5cm

\centerline{Stephen Bruce Sontz\footnote{Research partially
supported by CONACYT (Mexico) project 49187.}}

\centerline{Centro de Investigaci\'on en Matem\'aticas, A.C. (CIMAT)}

\centerline{Guanajuato, Mexico}

\centerline{email: sontz@cimat.mx}

\vskip .8cm

\begin{abstract}
\noindent
This note explains how the two measures
used to define the $\mu$-deformed Segal-Bargmann space
are natural and essentially unique structures.
As is well known,
the density with respect to Lebesgue measure of each of these measures
involves a Macdonald function.
Our primary result is that these densities are the solution of a system of
ordinary differential equations which is naturally associated with this theory.
We then solve this system and find the known densities as well as
a ``spurious'' solution which only leads to a trivial holomorphic Hilbert space.
This explains how the Macdonald functions arise in this theory.
Also we comment on why it is plausible that only one measure will not work.
We follow Bargmann's approach by imposing a condition
sufficient for the $\mu$-deformed creation and annihilation operators
to be adjoints of each other.
While this note uses elementary techniques, it reveals in a new way basic aspects of the
structure of the $\mu$-deformed Segal-Bargmann space.
\end{abstract}
\vskip .3cm

\section{Introduction}
\label{sec1}

Before getting into details, let us remark that we will
be studying deformations, depending on a dimensionless deformation
parameter $\mu > -1/2$ (which is fixed throughout the note),
of standard analysis and quantum mechanics.
Our goal is to provide motivation for Definitions \ref{defmumeas}
and \ref{defmusbs} below, including an explanation of
the appearance of the Macdonald function in those definitions.

   We start by recalling some definitions and notation that appear in \cite{CAASBS}.
Also see \cite{CAA}, \cite{PS}, \cite{PS2}, \cite{PS3} and \cite{SBS} and references therein
for other related work.

\begin{definition}
\label{defmumeas}
Say $\lambda > 0$
We define measures on the complex plane  ${\bf C}$ by
\begin{gather*}
d\nu_{e,\mu,\lambda}(z) := \nu_{e,\mu,\lambda}(z) dx dy, \\
d\nu_{o,\mu,\lambda}(z) := \nu_{o,\mu,\lambda}(z) dx dy,
\end{gather*}
whose densities are defined by
\begin{gather}
\label{defdene}
\nu_{e,\mu,\lambda}(z) := \lambda 
    \frac{2^{ \frac{1}{2}-\mu }}{\pi \Gamma(\mu+\frac{1}{2} )}
 K_{\mu-\frac{1}{2}} ( | \lambda^\frac{1}{2} z|^2 )
| \lambda^\frac{1}{2} z|^{2\mu +1}  \ , \\
\label{defdeno}
\nu_{o,\mu,\lambda}(z) :=
\lambda \frac{2^{ \frac{1}{2}-\mu }}{\pi \Gamma(\mu+\frac{1}{2} )}
  K_{\mu+\frac{1}{2}} ( | \lambda^\frac{1}{2} z|^2 )
| \lambda^\frac{1}{2} z|^{2\mu +1}
\end{gather}
for $0 \ne z \in {\bf C}$,
where $\Gamma$  (the Euler gamma function) and $K_\alpha$  (the Macdonald function
of order $\alpha$) are defined in \cite{LEB}.
Moreover, $dxdy$ is Lebesgue measure on ${\bf C}$.
\end{definition}

    The function $K_\alpha$ is also known as the modified
Bessel function of the third kind or Basset's function.
(See \cite{ER}, p.~5.)
But it is also simply known as a modified Bessel function.
(See \cite{GRAD}, p.~961, and \cite{AB}, p. 374.)
An explanation of where the Macdonald functions in Definition \ref{defmumeas}
come from was the motivation for writing this note.

    From the formulas (\ref{defdene}) and (\ref{defdeno}),
one can see why the case $\mu = -1/2$ has not been included.
One should refer to the discussion of the Bose-like oscillator in \cite{RO} (especially,
note Theorem~5.7) for motivation for the condition $\mu > -1/2$.

    Let ${\cal H} ({\bf C}) $ be the space of all holomorphic functions
$f : {\bf C} \rightarrow {\bf C}$.
We note that $f_e := (f + Jf)/2$ (respectively, $f_o := (f - Jf)/2$) defines the even (respectively, odd) part of
$f$, where $Jf(z):= f(-z)$ for all $z\in {\bf C}$ is the parity operator.
So, $ f = f_e + f_o$.

    We use throughout the article the standard notations for $L^2$ spaces,
for their inner products, and for their norms.

\begin{definition}
\label{defmusbs}
The \emph{$\mu$-deformed Segal-Bargmann space} for $\lambda > 0$  is
\begin{equation*}
   {\cal B}^2_{\mu,1/\lambda} :=
   {\cal H} ({\bf C})  \cap \left\{ f : {\bf C} \rightarrow
   {\bf C} \ | \  f_e \in L^2( {\bf C}, \nu_{e,\mu,\lambda})
   {\rm ~and~}   f_o \in L^2( {\bf C}, \nu_{o,\mu,\lambda}) \right\},
\end{equation*}
where $ f = f_e + f_o$ is the decomposition of a function into its even and odd parts.
Next we define the norm
$$ 
   || f ||_{ {\cal B}^2_{\mu,1/\lambda} } :=
    \left(  || f_e ||^2_{ L^2( {\bf C}, \nu_{e,\mu,\lambda} )}
            + || f_o ||^2_{ L^2( {\bf C}, \nu_{o,\mu,\lambda} )}
                                                     \right)^{1/2}
$$
for all $ f \in {\cal B}^2_{\mu,1/\lambda}$.
\end{definition}

   This definition is due to Marron in \cite{MA} and Rosenblum in \cite{RO2}.
The reason for using $1/\lambda$ instead of $\lambda$ in the notation
has to do with maintaining consistency with the notation of Hall in \cite{HA1}.
For more on the historical background of this definition, see \cite{CAASBS}.
We have that ${\cal B}^2_{\mu,1/\lambda}$
is a Hilbert space (see \cite{MA}) with inner product defined by
\begin{equation} 
\label{defip}
\langle f, g \rangle_{ {\cal B}^2_{\mu,1/\lambda} } 
:= \langle f_e, g_e \rangle_{L^2(\nu_{e,\mu,\lambda})} + \langle f_o, g_o \rangle_{L^2(\nu_{o,\mu,\lambda})}.
\end{equation}
Of course, $f = f_e + f_o$ and $g = g_e + g_o$ are the representations of $f$ and $g$
as the sums of their even and odd parts.
(We will often use such representations without explicit comment, letting the notation 
carry the burden of explanation.)
When $\mu = 0$ and $\lambda=1$
this reduces to the usual Segal-Bargmann space, denoted here by $ {\cal B}^2$.
(See \cite{BA, SEG}.)
For simplicity of notation we put $\lambda = 1$ for the rest of this article.
We also put $\mathcal{B}^2_\mu = \mathcal{B}^2_{\mu ,1}$

   There is a way of seeing a relation of this article to standard Segal-Bargmann analysis.
In general this is to find relations in the $\mu$-deformed theory that do not
depend on the parameter $\mu$.
This means that the relations for $\mu \ne 0$ are exactly the same as those
for the standard case $\mu=0$.
This will be our approach in Section~\ref{sec2} where we motivate the definition
of the measures in the $\mu$-deformed Segal-Bargmann space and, in particular,
show how the Macdonald functions arise naturally.

\section{The Measures in the Segal-Bargmann spaces}
\label{sec2}

It turns out that $\mathcal{B}^2_\mu$ is the image of a $\mu$-deformed Segal-Bargmann transform
(see \cite{MA}) which is unitary.
This unitarity is a consequence of the definition of the inner product on $\mathcal{B}^2_\mu$
in terms of the two measures defined in Definition 1.1.
We wish to motivate this definition of the inner product intrinsically, that is, without reference
to the $\mu$-deformed Segal-Bargmann transform but rather as a basic structure that
arises naturally for holomorphic functions.
We will do this modulo a normalization factor that is left undetermined intrinsically.
Rather than prove theorems, the purpose of this note is
to show how the definitions (\ref{defdene})
and (\ref{defdeno}) are naturally motivated.
We follow an idea given in \cite{BA}.
To achieve this we will use $\mu$-deformed creation and annihilation operators
defined for arbitrary holomorphic functions.
\begin{definition}
Let $f \in {\cal H}({\bf C })$ be a holomorphic function and $z \in {\bf C}$.
Then the $\mu$-deformed creation operator
$a^*_\mu : {\cal H}({\bf C }) \to {\cal H}({\bf C })$ is defined by
$$
    a^*_\mu f (z) := z f(z).
$$
The $\mu$-deformed annihilation operator
$a_\mu : {\cal H}({\bf C }) \to {\cal H}({\bf C })$ is defined by
$$
   a_\mu f (z) := \dfrac{\partial f}{ \partial z } + \frac{\mu}{z}\Big( f(z) - f(-z) \Big),
$$
(Here we use the standard notation
$\partial /\partial z = (1/2) ( \partial /\partial x - i \partial /\partial y)$
from complex variable theory for the complex derivative operator.
We also mention in passing that $a_{\mu}$ is a (complex) Dunkl operator.)
\end{definition}

   These operators satisfy the $\mu$-deformed commutation relation
\begin{equation}
\label{holmuccr}
 [ a_\mu, a^*_\mu] = I + 2 \mu J
\end{equation}
on ${\cal H}({\bf C })$.
This commutation relation, which is the central identity of $\mu$-deformed quantum mechanics,
was originally introduced by Wigner in \cite{WI}.

   For example, in the original theory in \cite{BA} when $\mu = 0$, Bargmann defines
the inner product in terms of a measure on the phase space, namely,
$$
  \left\langle  f, g \right\rangle_{ {\cal B}^2_0 } :=
\int_{ {\bf C} } dx dy \, \nu_{{\rm Gauss}}(z)  f(z)^* g(z)
$$
for all holomorphic functions $f$ and $g$ that are in
$L^2 ( {\bf C}, \nu_{ {\rm Gauss}} )$.
Here, the density $\nu_{{\rm Gauss}}$ is defined by
\begin{equation}
         \nu_{ {\rm Gauss}} (z) := \frac{1}{\pi} e^{- |z|^2}
\end{equation}
for all $z \in \mathbf{C}$.
Moreover, Bargmann's motivation for this definition of the measure on the phase space
comes from formally analyzing
for \emph{holomorphic} $f$ and $g$ the condition
$$
 \left\langle a^*_0 f, g \right\rangle_{ L^2  (  {\bf C}, \nu) } =
 \left\langle  f,  a_0 g \right\rangle_{ L^2  (  {\bf C}, \nu) },
$$
where $\nu(z)$ is an unknown density function which defines
the measure $\nu(z) dx dy $ on ${\bf C}$.
We can try a similar strategy for the $\mu$-deformed
creation and annihilation operators $a^*_\mu$ and $a_\mu$ in place of
the usual creation and annihilation operators $a^*_0$ and $a_0$.
So we want to consider, again for $f$ and $g$ holomorphic, the identity
$$
 \left\langle a^*_\mu f, g \right\rangle_{ L^2  (  {\bf C}, \nu_\mu) } =
 \left\langle  f,  a_\mu g \right\rangle_{ L^2  (  {\bf C}, \nu_\mu) }
$$
and try to find the unknown density function $\nu_\mu(z)$ for
a measure $\nu_\mu(z) dx dy $ on ${\bf C}$.
We have been unable to prove that this has no solution when $\mu \ne 0$,
though this seems to be the case.
However, we have shown that the sufficient condition on $\nu_\mu(z)$ given
by a formal integration by parts argument (as done by Bargmann in \cite{BA})
has no solution if $\mu \ne 0$.
We will come back to this point later.

      In any event, what Marron in \cite{MA} and
Rosenblum in \cite{RO2} did was to define \emph{two} measures on the
phase space ${\bf C}$ and use formula (\ref{defip})
to define the inner product.
We do not know what motivation they had to write down these measures, but we have been able
to construct the following intuitive reasoning \emph{\`a la} Bargmann in \cite{BA}.
We do not believe that the following exposition is new.
Indeed, we fully expect it was known to Rosenblum.
However, we have not found it in the literature.
Unfortunately, Marvin Rosenblum died some time after giving us a copy of
\cite{RO2}, which is a sketchy preliminary document that was as far
as we know never put into a publishable form.

   First, we consider the desired relation
\begin{equation}
\label{desirelation}
 \left\langle a^*_\mu f, g \right\rangle_{ {\cal B}^2_\mu } =
 \left\langle  f,  a_\mu g \right\rangle_{ {\cal B}^2_\mu }.
\end{equation}

  Since the non-local parity operator $J$ figures in the $\mu$-deformed canonical
commutation relation (\ref{holmuccr}),
it seems plausible to divide ${\cal H} ( {\bf C} )$ into the two eigenspaces for
this operator, i.e., the subspaces of even and odd functions, respectively.
So we propose to introduce two measures with densities $\nu_e$ and $\nu_o$
on the phase space ${\bf C}$ and define an inner product by
\begin{equation}
\label{newipdef}
\left\langle  f,  g \right\rangle_{ {\cal B}^2_\mu } := \left\langle  f_e,  g_e \right\rangle_{ L^2(\nu_e) } +
              \left\langle  f_o,  g_o \right\rangle_{ L^2(\nu_o)  },
\end{equation} 
using the even and odd parts of $f$ and $g$ on the right hand side of this definition.

   Next, we want to see what restriction (\ref{desirelation}) places on the unknowns
$\nu_e$ and $\nu_o$.
Since $a^*_\mu$ and $a_\mu$ interchange the even and odd subspaces
of ${\cal H} ( {\bf C} )$
and since these are orthogonal subspaces for the proposed inner product (\ref{newipdef}),
there are exactly two non-trivial cases of (\ref{desirelation}).
The first such case is for $f$ even and $g$ odd.
Then we have the condition
\begin{equation}
\label{condition1}
 \left\langle a^*_\mu f, g \right\rangle_{ L^2(\nu_o) } =
 \left\langle  f,  a_\mu g \right\rangle_{ L^2(\nu_e) }.
\end{equation}
The second non-trivial case is for $f$ odd and $g$ even, in which case we have 
\begin{equation}
\label{condition2}
 \left\langle a^*_\mu f, g \right\rangle_{ L^2(\nu_e) } =
 \left\langle  f,  a_\mu g \right\rangle_{ L^2(\nu_o) }.
\end{equation}

    Next, we write out these conditions in terms of integrals.
The first condition (\ref{condition1}) gives
\begin{gather}
 \int_{ {\bf C} } dx dy \, \nu_o(z) z^* f(z)^* g(z) = 
       \int_{ {\bf C} } dx dy \, \nu_e(z) f(z)^* a_\mu g(z) \nonumber \\
 = \int_{ {\bf C} } dx dy \, \nu_e(z) f(z)^*
    \left( \frac{\partial g}{\partial z} + \frac{2\mu}{z} g(z) \right) \nonumber \\
 = \int_{ {\bf C} } dx dy \left( - \frac{\partial \nu_e}{\partial z} +
    \nu_e(z) \frac{2\mu}{z} \right) f(z)^* g(z),
\label{intcond1}
\end{gather}
where we used the fact that $g$ is odd to calculate $a_\mu g$ and then we integrated
formally by parts,
using $ \partial f^* / \partial z = 0$ since $f$ is holomorphic.

   The second condition (\ref{condition2}) gives
\begin{gather}
 \int_{ {\bf C} } dx dy \, \nu_e(z) z^* f(z)^* g(z) = \int_{ {\bf C} } dx dy \, 
     \nu_o(z) f(z)^* a_\mu g(z) \nonumber \\
 = \int_{ {\bf C} } dx dy \, \nu_o(z) f(z)^*  \frac{\partial g}{\partial z} \nonumber \\
 = \int_{ {\bf C} } dx dy \left( - \frac{\partial \nu_o}{\partial z} \right)  f(z)^* g(z),
\label{intcond2}
\end{gather}
where we used the fact that $g$ is even to evaluate $a_\mu g$.
Again, the last equality is a formal integration by parts.
Clearly a \emph{sufficient} condition for these two conditions (\ref{intcond1}) and
(\ref{intcond2})
to hold is this system:
\begin{gather*}
      z^*   \nu_o(z)  = - \frac{\partial \nu_e}{\partial z} + \nu_e(z) \frac{2\mu}{z}, \\
      z^*   \nu_e(z)  = - \frac{\partial \nu_o}{\partial z},
\end{gather*}
which is equivalent to
\begin{gather}
\label{sys1}
      |z|^2   \nu_o(z)  = - z \frac{\partial \nu_e}{\partial z} + 2 \mu \nu_e(z), \\
\label{sys2}
      |z|^2   \nu_e(z)  =  - z \frac{\partial \nu_o}{\partial z}.
\end{gather}
While this is a sufficient condition for
(\ref{condition1}) and (\ref{condition2}),
it is not clear whether it is also necessary, since the
functions $f$ and $g$ in the integral identities
(\ref{condition1}) and (\ref{condition2}) are both holomorphic and of specific parities.

   If we try to use only one measure on $\mathbf{C}$ and impose Bargmann's condition, we find (by almost
the same argument) that (\ref{sys1}) and (\ref{sys2}) hold provided that we set
$\nu_e = \nu_o$ in them.
But this pair of equations would then have no nonzero solution for $\mu \ne 0$.
This leads us to believe that the $\mu$-deformed Segal-Bargmann space can not be
realized as a subspace of $L^2(\mathbf{C}, \nu)$ for some measure $\nu$ that is
absolutely continuous with respect to Lebesgue measure.

   Next we note that in the standard polar coordinates $r$, $\theta$
in ${\bf C}$ we have that
$$
 z \frac{\partial}{\partial z} =  \frac{1}{2}
\left( r  \frac{\partial}{\partial r} - i  \frac{\partial}{\partial \theta} \right).
$$
Since we are seeking \emph{real} solutions $\nu_e$ and $\nu_o$, it follows by equating
imaginary (respectively, real) parts on both sides of (\ref{sys1}) and (\ref{sys2}),
when written in polar coordinates, that
\begin{gather*}
          \frac{\partial}{\partial \theta} \nu_e = 0, \\
          \frac{\partial}{\partial \theta} \nu_o = 0, \\
          r^2 \nu_o = - \frac{1}{2} r \frac{\partial}{\partial r} \nu_e + 2 \mu \nu_e, \\
          r^2 \nu_e = - \frac{1}{2} r \frac{\partial}{\partial r} \nu_o.
\end{gather*}
Thus, we are looking for two functions of only $r>0$ (since they do not
depend in $\theta$ according to the first two equations) satisfying these last two equations, which
form a first order, homogeneous linear system with variable real-valued
coefficients in the unknown pair ($\nu_e,\nu_o)$.
Therefore these real-valued solutions satisfy the coupled equations
\begin{gather}
\label{couple1}
  \nu_o = - \frac{1}{2r}  \frac{d}{d r} \nu_e + \frac{2 \mu}{r^2} \nu_e, \\
\label{couple2}
  \nu_e = - \frac{1}{2r}  \frac{d}{d r} \nu_o.
\end{gather}
and so form a vector space of dimension two over ${\bf R}$.
Substituting the first equation into the second, we get
\begin{equation}
\label{sepnue}
\frac{d^2 \nu_e}{d r^2} - \frac{(1+4\mu)}{r}
\frac{d \nu_e}{d r} + \left( \frac{8\mu}{r^2} - 4r^2 \right) \nu_e =0,
\end{equation}
while by substituting the second into the first, we have that
\begin{equation}
\label{sepnuo}
\frac{d^2 \nu_o}{d r^2} - \frac{(1+4\mu)}{r}
\frac{d \nu_o}{d r} - 4r^2 \nu_o =0.
\end{equation}
Now (\ref{sepnue}) and (\ref{sepnuo}) are decoupled second order linear differential
equations, which impose sufficient conditions on the solutions $\nu_e$ and $\nu_o$
of the original coupled system (\ref{couple1}) and (\ref{couple2}).
Since each one of the equations (\ref{sepnue}) and (\ref{sepnuo}) has a space
of real-valued solutions of dimension two over $ {\bf R}$, the pairs of these
solutions ($\nu_e,\nu_o)$ form a space of dimension four, which includes the two
dimensional space of solutions of the coupled system (\ref{couple1}) and (\ref{couple2}).
Our method will be to find the general solution of both of the equations
(\ref{sepnue}) and (\ref{sepnuo})
and then identify the two dimensional subspace of solutions to the coupled system.

   Making the change of dependent variable,
$\nu_e(r)= r^\alpha \phi(r^2)$ in (\ref{sepnue})
we find that
\begin{equation}
\label{eq4phi}
    \phi^{\prime \prime} (r^2) + \frac{\alpha - 2 \mu}{r^2} \phi^\prime (r^2)
+ \left(
       \frac{\alpha^2 -2\alpha -4 \alpha \mu + 8 \mu}{4r^4} - 1
  \right)
\phi(r^2) =0,
\end{equation}
which looks something like Bessel's equation of order $\nu$ (\cite{LEB}, p.~98), namely
$$
     u^{\prime \prime}(x) + \dfrac{1}{x} u^{\prime}(x) + \left( 1 - \dfrac{\nu^2}{x^2} \right) u(x) = 0
$$
for $x = r^2$.
(This change of the variable may seem unmotivated, as is often the case
with this method, but it is really not that unusual.)
To get better agreement with the form of Bessel's equation, we choose the exponent
$\alpha$ in the change of variable
such that $\alpha - 2\mu = 1$, that is $\alpha = 2\mu +1$.
With this value for $\alpha$ we calculate that
$\alpha^2 -2\alpha -4 \alpha \mu + 8 \mu = -4(\mu -1/2)^2 $ and so
(\ref{eq4phi}) becomes
\begin{equation}
\label{cdv1}
    \phi^{\prime \prime} (r^2) + \frac{1}{r^2} \phi^\prime (r^2)
- \left(
       \frac{(\mu -1/2)^2}{r^4} + 1
  \right)
\phi(r^2) =0,
\end{equation}
which is not Bessel's equation but rather a related
equation known as Bessel's modified equation  of order $\nu$, namely
$$
   u^{\prime \prime}(x) + \dfrac{1}{x} u^{\prime}(x) - \left( 1 + \dfrac{\nu^2}{x^2} \right) u(x) = 0
$$
with $x=r^2$ and $\nu = \mu - 1/2$.
Its general real valued solution (\cite{LEB}, p.~110) is
$$
\phi(r^2) = a I_{\mu-1/2}(r^2) +b K_{\mu-1/2}(r^2),
$$
where
$I_{\mu-1/2}$ is the modified Bessel function of the first kind and of order $\mu-1/2$,
$K_{\mu-1/2}$ is the Macdonald function of order $\mu-1/2$
and $a,b \in {\bf R}$.
See \cite{LEB} for more details about these special functions.
Consequently,
$$
\nu_e(r) = r^{2\mu +1} [ a I_{\mu-1/2}(r^2) +b K_{\mu-1/2}(r^2)]
$$
is the general real valued solution of (\ref{sepnue}), where $a,b \in \mathbf{R}$.

   Similarly the change of variable, $\nu_o(r)= r^\alpha \psi(r^2)$ converts (\ref{sepnuo}) into
\begin{equation}
\label{eq4psi}
    \psi^{\prime \prime} (r^2) + \frac{\alpha - 2 \mu}{r^2} \psi^\prime (r^2)
+ \left(
       \frac{\alpha^2 -2\alpha -4 \alpha \mu }{4r^4} - 1
  \right)
\psi(r^2) =0,
\end{equation}
and again we choose $\alpha = 2 \mu +1$ for the same reason as before.
So, $\alpha^2 -2\alpha -4 \alpha \mu  = -4(\mu + 1/2)^2 $ follows
and (\ref{eq4psi}) becomes
\begin{equation}
    \psi^{\prime \prime} (r^2) + \frac{1}{r^2} \psi^\prime (r^2)
- \left(
       \frac{(\mu + 1/2)^2}{r^4} + 1
  \right)
\psi(r^2) =0.
\end{equation}
This is again Bessel's modified equation, but now of order
$\mu + 1/2$ instead of $\mu - 1/2$.
So its general real valued solution is
$$
   \psi(r^2) = c I_{\mu+1/2}(r^2) +d K_{\mu+1/2}(r^2),
$$
where $c,d \in {\bf R}$.
Thus
$$
\nu_o(r) = r^{2\mu +1} [ c I_{\mu+1/2}(r^2) +d K_{\mu+1/2}(r^2)]
$$
is the general real valued solution of (\ref{sepnuo}).
It only remains to eliminate the superfluous solutions, namely the solutions of
the individually decoupled equations that do not pair up to give a solution of
the coupled system (\ref{couple1}) and (\ref{couple2}).
For example, starting with the right side of (\ref{couple2}) and putting
$\nu_o(r) = r^{2\mu +1} K_{\mu+1/2}(r^2)$ and using $s=r^2$ we see that
\begin{gather*}
\left( -\frac{1}{2r} \right) \frac{d}{d r}
\left( r^{2\mu+1} K_{\mu+1/2} (r^2) \right)
= - \frac{d}{d s} \left( s^{\mu+1/2} K_{\mu+1/2}(s) \right) \\
= s^{\mu+1/2} K_{\mu-1/2}(s) = r^{2\mu+1} K_{\mu-1/2}(r^2)
\end{gather*}
where the second equality is an identity that can be found in \cite{LEB}, p.~110.
This shows that the pair
\begin{equation}
\label{pairk}
(r^{2\mu +1} K_{\mu-1/2}(r^2), r^{2\mu +1} K_{\mu+1/2}(r^2))
\end{equation}
is a solution of the coupled system.
Since $K_{ - 1/2 }(z) = K_{ 1/2 }(z)$ (see \cite{LEB}),
we see that these two densities are equal when $\mu =0$.
But when $\mu \ne 0$, these densities are {\em not} equal and so we find that
our sufficient condition does not give one measure on the phase space, but rather two.

    Since the coupled system solution space has dimension two, we
still need one more linearly independent pair solving the coupled system.
But we have the following calculation that is very similar to the previous one:
\begin{gather*}
\left( -\frac{1}{2r} \right) \frac{d}{d r}
\left( r^{2\mu+1} I_{\mu+1/2} (r^2) \right)
= - \frac{d}{d s} \left( s^{\mu+1/2} I_{\mu+1/2}(s) \right) \\
= - s^{\mu+1/2} I_{\mu-1/2}(s) = - r^{2\mu+1} I_{\mu-1/2}(r^2)
\end{gather*}
by another identity from \cite{LEB}, p.~110.
So the pair
\begin{equation}
\label{pairi}
( -r^{2\mu +1} I_{\mu-1/2}(r^2), r^{2\mu +1} I_{\mu+1/2}(r^2) )
\end{equation}
is another, linearly independent solution of the coupled system.
Now, $K_\nu(x) > 0$ and $I_\nu(x) > 0$ for all $x > 0$.
This means that the solution (\ref{pairk}) gives a pair of
positive densities, which then define positive measures.
However, the solution (\ref{pairi}) gives a pair of densities
with opposite signs even after multiplying by any nonzero real number.
This in itself is not a fatal flaw with the solution (\ref{pairi}) since
one can develop interesting theories with signed measures.
The real problem with this solution is that its asymptotic behavior when
$r \to \infty$ is growth in absolute value to infinity for each function in the pair.
(Precise information on this growth can be found in \cite{LEB} but that
is not relevant to this discussion.)
Since we are looking for measures for {\em holomorphic} functions which, except for
the constants, also go to infinity for $r \to \infty$, there is no way to use the
solution (\ref{pairi}) to construct a pair of nontrivial holomorphic $L^2$ function spaces.
On the other hand, both of the
functions in (\ref{pairk}) are integrable with respect to the measure $r dr$,
which is the radial part of Lebesgue measure $ r dr d\theta$ in polar coordinates.
(See \cite{CAASBS} or \cite{PS2}.)
With a suitable normalization either one of the functions in (\ref{pairk})
(but not both if $\mu \ne 0$) can be made into a probability measure.
The definition in (\ref{defdene}) and (\ref{defdeno}) has the normalization that
makes $d\nu_{e,\mu}$ into a probability measure.
(Recall again that we are only considering in detail here the case $\lambda =1$, but
the case for general $\lambda > 0$ follows immediately.)

   We can understand the particular normalization in (\ref{defdene}) and (\ref{defdeno}) in terms of the
$\mu$-deformed Segal-Bargmann transform $B_{\mu}$.
(This paragraph uses definitions from \cite{SBS}, which should be consulted
for more details.)
It turns out that $B_{\mu}$ has been normalized so that $B_{\mu} 1 = 1$.
But $1$ is a unit vector in the domain $L^2 ( {\bf R}, d\rho_{\mu} ) $ of $B_\mu$,
since $d\rho_{\mu}$ is a probability measure.
Since we want $ B_{\mu}$ to be a unitary transform, a necessary condition is that $1$ in the
codomain also be a unit vector.
Since $1$ is an even function, this forces $d\nu_{e,\mu}$ to be a probability measure,
while imposing no restriction on $d\nu_{o,\mu}$.
Clearly the normalization of the two measures can not be made intrinsic
to the holomorphic side of the theory by
just using the Bargmann condition, which is itself homogeneous in the measures.

    We would like to close this presentation of our understanding of
where the Macdonald functions in the $\mu$-deformed Segal-Bargmann space
come from by reiterating that all of
this discussion fits into the way of understanding
why this theory should be thought of as a type of Segal-Bargmann analysis,
as we described at the very end of Section~\ref{sec1}.
This is because the relation between the creation and annihilation operators,
namely that they are adjoints, does not depend on the deformation
parameter $\mu$.

\section{Open Problems}
   A completely different way was introduced by Hall in \cite{HA1} for
defining an intrinsic inner product
on the codomain of his generalized Segal-Bargmann transforms.
This is done in terms of a heat kernel measure defined on the phase space.
We do not go into details here, but merely note that his method produces only
one measure (for each of his three versions: $A$, $B$ and $C$), and so it appears
not to be applicable to the case of the $\mu$-deformed Segal-Bargmann transform,
where we have \emph{two} measures when $ \mu \ne 0$.
Notice that this relates to the general problem posed in \cite{SBSBO} 
of finding some measure  $\nu$ on the phase space in order to realize an abstract Hilbert space, introduced there as a generalized Segal-Bargmann space associated with a Coxeter group, as a subspace of $L^2(\nu)$.
In particular, it indicates that the problem may be to find a finite
family of measures on the phase space instead of merely one.
It would be interesting if one could find a construction of
the measures for the $\mu$-deformed Segal-Bargmann space using
methods from heat kernel analysis, although this can not be in strict
analogy with Hall's method as we have noted above.
We consider this is be a major challenge remaining in this
area of research.

Nothing in our discussion excludes the possibility
 that there may well be ways of putting three or more
measures on the phase space and using them to construct a Segal-Bargmann
type space together with an associated Segal-Bargmann transform.
And we have not proved definitively that this theory can not be made to
work with only one measure, though this seems plausible on account of our earlier remarks.
In this context, we should note that Asai has shown in \cite{AS}
under some rather stringent hypotheses
that the Segal-Bargmann space associated to a probability measure
on the configuration space $\mathbf{R}$
can be realized as the $L^2$ space of holomorphic functions on the phase space $\mathbf{C}$
for a unique probability measure on $\mathbf{C}$.
However, the case of $\mu$-deformed quantum mechanics considered here is
not included among the cases considered in \cite{AS}.

   Finally, the ``spurious'' solution which we have found
for the coupled system (\ref{couple1}) and (\ref{couple2}) may still be useful in
the construction of some sort of related theory. This is a highly speculative
as well as  vague comment.

\section{Acknowledgments}
This note was begun during an academic visit at the Department of Mathematics
of the University of Virginia during my sabbatical year in 2007.
I wish to express my thanks to Larry Thomas, my host there.
I also wish to thank N.~Asai for bringing his article \cite{AS} to my attention.
And I thank all of the organizers of the 11th Workshop on Non-harmonic Analysis
held in Bedlewo, Poland, during the week of 18-23 August 2008 for their cordial invitation
to participate in that intellectually interesting, and challenging, event.

\end{document}